\documentclass[twoside]{ae100prg}
\usepackage{amsmath}
\usepackage{graphicx}
\usepackage[breaklinks]{hyperref}
\usepackage{booktabs}

\input{tcilatex}
\begin{document}

\title{Gravitational waveforms for black hole binaries with unequal masses}
\author{M\'{a}rton T\'{a}pai$^{1}$, Zolt\'{a}n Keresztes$^{1}$, L\'{a}szl%
\'{o} \'{A}rp\'{a}d Gergely$^{1}$}
\email{tapai@titan.physx.u-szeged.hu}

\begin{abstract}
We derived a post-Newtonian (PN) inspiral only gravitational waveform for
unequal mass, spinning black hole binaries. Towards the end of the inspiral
the larger spin dominates over the orbital angular momentum (while the
smaller spin is negligible), hence the name Spin-Dominated Waveforms (SDW).
Such systems are common sources for future gravitational wave detectors and
during the inspiral the largest amplitude waves are emitted exactly in its
last part. The SDW waveforms emerge as a double expansion in the PN
parameter and the ratio of the orbital angular momentum to the dominant spin.
\end{abstract}

\address{$^1$ University of Szeged, Departments of Theoretical and Experimental Physics, D\'{o}m t\'{e}r 9. 6720 Szeged, Hungary}

\section{Introduction}

Gravitational wave detectors like the Advanced LIGO (aLIGO), or the planned
Einstein Telescope (ET), LAGRANGE and eLISA (NGO) space missions will
measure gravitational waves from black hole binaries of various total masses 
$m$. For astrophysical black hole binaries (with total mass $m$ a few ten
times of the mass of the sun M$_{\odot }$), the comparable mass and the
unequal mass case are both likely. For supermassive black hole binaries
(total mass is between $10^{6}~$M$_{\odot }~\div 10^{10}~$M$_{\odot }$) the
typical mass ratio $\nu $ is between $0.3$ and $0.03$ \cite{spinflip1}, \cite%
{spinflip2}.

For unequal masses the mass ratio can stand as a second small parameter. The
purpose of this paper is to give an approximation for the gravitational
waveforms in the small mass ratio regime.

\section{Spin-dominated waveforms}

It was shown in Ref. \cite{spinflip1}, that for rapidly spinning black hole
binaries, the smaller spin is of order $\nu ^{2}$ compared to the dominant
spin $S_{1}$, thus it can be neglected to first order in $\nu $. Furthermore
the ratio of the orbital angular momentum $L_{N}$ and $S_{1}$ was also given 
\cite{spinflip1}%
\begin{equation}
\frac{S_{1}}{L_{N}}\approx \varepsilon ^{1/2}\nu ^{-1}\chi _{1~,}
\label{sperl}
\end{equation}%
where $\varepsilon =Gm/c^{2}r\approx v^{2}/c^{2}$ (with $r$ the orbital
separation and$\ v$ the orbital velocity of the reduced mass particle $\mu
=m_{1}m_{2}/m$, $G$ the gravitational constant, $c$ the speed of light) is
the post-Newtonian (PN) parameter and $\chi _{1}\in \left[ 0,1\right] $ is
the dimensionless spin. For maximally spinning black holes $\chi _{1}=$ $1$.

As the PN parameter increases throughout the inspiral, the relation (\ref%
{sperl})\ shows, that $S_{1}$ will dominate over $L_{N}$ at the end of the
inspiral (thus the approximated waveforms are called Spin-Dominated
Waveforms, SDW). This condition at the technical level is included in the
smallness of the parameter $\xi =\varepsilon ^{-1/2}\nu $. 
\begin{table}[bbp]
\caption{SDW contributions of different $\protect\xi $ and $\protect%
\varepsilon $ orders. The SO terms contain the dominant spin.}
\label{table01}
\begin{center}
\begin{tabular}{l|llll}
\hline
& $\varepsilon ^{0}$ & $\varepsilon ^{1/2}$ & $\varepsilon ^{1}$ & $%
\varepsilon ^{3/2}$ \\ \hline
$\xi ^{0}$ & $h_{_{\times }^{+}}^{0}$ & $h_{_{\times }^{+}}^{0.5}$ & $%
h_{_{\times }^{+}}^{1},h_{_{\times }^{+}}^{1SO}$ & $h_{_{\times
}^{+}}^{1.5},h_{_{\times }^{+}}^{1.5SO},h_{_{\times }^{+}}^{1.5tail}$ \\ 
$\xi ^{1}$ & $h_{_{\times }^{+}}^{0\beta }$ & $h_{_{\times }^{+}}^{0.5\beta
} $ & $h_{_{\times }^{+}}^{1\beta },h_{_{\times }^{+}}^{1\beta SO}$ &  \\ 
\hline
\end{tabular}%
\end{center}
\end{table}

PN waveforms were previously calculated to 1.5 PN order \cite{kidder}, \cite%
{ABFO}, and to 2 PN order in\ Ref. \cite{BFH}. In order to approximate the
waveforms in the small mass ratio regime, we expand the waveforms in both
parameters $\varepsilon $ and $\xi $. The waveforms have the following
structure \cite{SDW}: 
\begin{subequations}
\begin{eqnarray}
h_{_{\times }^{+}} &=&\frac{2G^{2}m^{2}\varepsilon ^{1/2}\xi }{c^{4}Dr}\left[
h_{_{\times }^{+}}^{0}+\beta _{1}h_{_{\times }^{+}}^{0\beta }+\varepsilon
^{1/2}\left( h_{_{\times }^{+}}^{0.5}+\beta _{1}h_{_{\times }^{+}}^{0.5\beta
}-2\xi h_{_{\times }^{+}}^{0}\right) \right.   \notag \\
&&+\varepsilon \left( h_{_{\times }^{+}}^{1}-4\xi h_{_{\times
}^{+}}^{0.5}+\beta _{1}h_{_{\times }^{+}}^{1\beta }+h_{_{\times
}^{+}}^{1SO}+\beta _{1}h_{_{\times }^{+}}^{1\beta SO}\right)   \notag \\
&&\left. +\varepsilon ^{3/2}\left( h_{_{\times }^{+}}^{1.5}+h_{_{\times
}^{+}}^{1.5SO}+h_{_{\times }^{+}}^{1.5tail}\right) \right] ~,
\end{eqnarray}%
$D$ being the luminosity distance to the source. The terms are of different $%
\varepsilon $ and $\xi $ orders, as indicated in Table \ref{table01},and are
given in detail in Ref. \cite{SDW}. The angle $\beta _{1}$ span by $\mathbf{J%
}$ and $\mathbf{S}_{1}$ is of order $\xi $ too \cite{SDW}.

\begin{figure}[bbp]
\begin{center}
\includegraphics{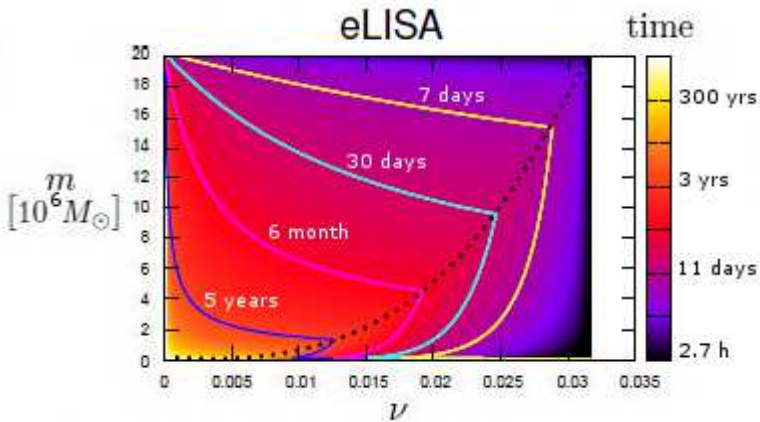}
\end{center}
\caption{The time interval $\Delta t$ until which the SDWs can be detected
by eLISA (NGO) as function of the total mass $m$ and mass ratio $\protect\nu 
$. $\Delta t$ either begins at the lower bound of the sensitivity range of
eLISA ($\protect\varepsilon _{f\min }$), or when the SDW approximation
begins to hold ($\protect\varepsilon _{1}$), and ends at the end of the
inspiral (chosen here as $\protect\varepsilon _{2}=0.1$). The color code is
logarithmic.}
\label{fig01}
\end{figure}

\section{Limits of validity}

We impose the smallness condition $\xi \leq 0.1$. This defines a lower limit of the
PN parameter $\varepsilon _{1}=Gm/c^{2}r_{1}=100\nu ^{2}$, implying an upper limit for the mass ratio, $\nu
_{\max }=0.0316\approx 1:32$. The upper limit for $\varepsilon $ is defined
by the end of the inspiral (chosen here as $\varepsilon _{2}=0.1$ \cite%
{LEVIN}).

From the expression $m=c^{3}\varepsilon ^{3/2}\left( \pi Gf\right) ^{-1}$
including the gravitational wave frequency $f$, also the leading order
radiative orbital angular frequency evolution \cite{omegadot} an integration
leads to the time $\Delta t$ during the binary evolves from $\varepsilon _{1}
$ to $\varepsilon _{2}$ 
\end{subequations}
\begin{equation}
\Delta t=\frac{5Gm}{2^{8}c^{3}}\frac{(1+\nu )^{2}}{\nu }\left( \varepsilon
_{1}^{-4}-\varepsilon _{2}^{-4}\right) ~.
\end{equation}
$\Delta t$ is shown as function of $m$ and $\nu $ on Fig \ref{fig01}. Even with the SDW approximation holding, the lower sensitivity bound ($f_{\min }=10^{-4}$ for
eLISA \cite{eLISA}) of the instrument may impose a larger value of the PN
parameter, as the lower validity bound $\varepsilon _{f_{\min }}$. Hence $%
\Delta t$ is calculated from $\max \left( \varepsilon _{1},\varepsilon
_{f_{\min }}\right) $ to $\varepsilon _{2}$.

A lower limit for the mass
ratio comes from the assumption that the second compact object has at least the
mass of a neutron star ($1.4~$M$_{\odot }$). The total mass is bounded from
above by the lower frequency bound of the detector (for eLISA $m=2\times
10^{7}$M$_{\odot }$, hence the minimal mass ratio for the eLISA detector is $%
\nu _{\min }=7\times 10^{-8}$).

\section{Concluding Remarks}

For unequal mass ratios the larger spin dominates over the orbital angular
momentum at the end of the inspiral. We have quantified this by the
introduction of a second small parameter $\xi $ and computed the respective
waveforms as a series expansion in both this and the PN parameter. A
comparison between the general waveforms of Ref. \cite{ABFO} and the SDWs
showed that the SDWs are approximately 80\% shorter, due to the smaller
parameter space and the second expansion in $\xi $. We expect the SDWs to be
useful tools in gravitational wave detection.

\section*{References}

\bibliographystyle{plain}
\bibliography{mtapai_prague}

\end{document}